\documentclass[twocolumn]{autart}    

\usepackage[utf8]{inputenc}
\usepackage{graphicx}         
\usepackage{amsfonts}
\usepackage{amssymb}  
\usepackage{amsmath,bm}
\usepackage{appendix}
\usepackage{natbib}
\usepackage{enumitem}
\usepackage{upgreek}
\usepackage[dvipsnames]{xcolor}

\usepackage[colorlinks=true,linkcolor={black},citecolor={black}]{hyperref}

\newtheorem{lemma}{Lemma}

\newcommand{\RR}{\mathbb{R}}

\newcommand{\x}{\mathbf{x}}
\newcommand{\nx}{x}
\newcommand{\nz}{{z}}
\newcommand{\nd}{{d}}

\newcommand{\toff}{t\sbrm{stop}}

\newcommand{\sbrm}[1]{\sb{\mathrm{#1}}}

\begin{document}

\begin{frontmatter}

\title{On inherent limitations in robustness and performance for a class of prescribed-time algorithms}

\thanks{Corresponding Author: R. Seeber}
\thanks[footnoteinfo2]{\textcolor{red}{This is the accepted manuscript version for:  Aldana-López R., Seeber R., Haimovich H., Gómez-Gutiérrez D. ``On inherent limitations in robustness and performance for a class of prescribed-time algorithms". Automatica. 2023, Article 111284; DOI: 10.1016/j.automatica.2023.111284. 
\textbf{Please cite the publisher's version}. For the publisher's version and full citation details, see:
\url{https://doi.org/10.1016/j.automatica.2023.111284}.
© 2023. This manuscript version is made available under the CC-BY-NC-ND 4.0 license \url{https://creativecommons.org/licenses/by-nc-nd/4.0/}}}
\thanks{Work partially supported by the Christian Doppler Research Association, the  Austrian Federal Ministry of Labour and
Economy and the National Foundation for Research, Technology and Development and by Agencia I+D+i grants PICT 2018-01385, 2021-0730, Argentina.}

\author[Zaragoza]{R.~Aldana-Lopez}\ead{rodrigo.aldana.lopez@gmail.com}, 
\author[Austria]{R.~Seeber}\ead{richard.seeber@tugraz.at},
\author[Argentina]{H.~Haimovich}\ead{haimovich@cifasis-conicet.gov.ar},
\author[Intel,TECMM]{D.~Gomez-Gutierrez}\ead{david.gomez.g@ieee.org}          

\address[Zaragoza]{University of Zaragoza, Departamento de Informatica e Ingenieria de Sistemas (DIIS), Zaragoza, Spain.}

\address[Austria]{Graz University of Technology, Institute of Automation and Control, Christian Doppler Laboratory for Model Based Control of Complex Test Bed Systems, Graz, Austria. }

\address[Argentina]{Centro Internacional Franco-Argentino de 
  Ciencias de la Informaci\'on y de Sistemas (CIFASIS)
  CONICET-UNR, 2000 Rosario, Argentina}
  
\address[Intel]{Intel Tecnolog\'ia de M\'exico, Intel Labs, Intelligent Systems Research Lab, Jalisco, Mexico.}

\address[TECMM]{Tecnol\'ogico Nacional de M\'exico, Instituto Tecnol\'ogico Jos\'e Mario Molina Pasquel y Henr\'iquez, Unidad Acad\'emica Zapopan, Jalisco, Mexico.}


\begin{abstract}                          %
Prescribed-time algorithms based on time-varying gains may have remarkable properties, such as regulation in a user-prescribed finite time that is the same for every nonzero initial condition and that holds even under matched disturbances. However, at the same time, such algorithms are known to lack robustness to measurement noise. This note shows that the lack of robustness of a class of prescribed-time algorithms is of an extreme form. Specifically, we show the existence of arbitrarily small measurement noises causing considerable deviations, divergence, and other detrimental consequences. We also discuss some drawbacks and trade-offs of existing 
workarounds as motivation for further analysis.
\end{abstract}

\end{frontmatter}

\section{Introduction}

Design methodologies that arbitrarily prescribe the convergence time bound of dynamical systems, such as closed control loops or observer error dynamics, have recently seen a great deal of attention.
Specifically, such prescribed-time algorithms achieve a so-called fixed convergence-time bound \citep[cf.][]{Polyakov2012a} that is arbitrarily prescribed and independent of the initial condition.
A subclass of these methodologies uses time-varying gains (TVG) that tend to infinity as the time approaches the prescribed convergence time. This is the case of ~\cite{Song2019Time-varyingTime,Song2017Time-varyingTime,Holloway2019,Orlov2022Prescribed-TimeGains,Aldana-Lopez2021AGains,Gomez2020RNC,Tran2020Finite-timeApproach} and \cite{Orlov2022TimeDesigns}. Compared to time-invariant approaches, such as \cite{Seeber2021RobustTime,Sanchez-Torres2018}, TVG-based approaches have been shown to have some remarkable advantages.
On the one hand, controllers for an integrator chain can be designed to achieve exact tracking \emph{at} a fixed, prescribed time that is the same for \emph{all trajectories}~\citep{Song2017Time-varyingTime,Song2019Time-varyingTime}, which maintain a prescribed-time convergence in the presence of bounded disturbances even without knowledge on its bound. Such methods can be extended for the output tracking problem for systems described by partial differential equations~\citep{Steeves2020Prescribed-timePDEs}. On the other hand, the observers by~\cite{Holloway2019} allow to reconstruct the system's state at a prescribed time instant, which can be extended to maintain the prescribed-time convergence property even under input delay~\citep{Espitia2022SensorSystems}.
\cite{Aldana-Lopez2021AGains} and \cite{Orlov2022Prescribed-TimeGains} designed online differentiators with a prescribed upper bound for the convergence time.

Despite the singularity of the TVG at the desired convergence time instant or convergence-time bound, the magnitude of controllers and the error correction functions of differentiators, in the absence of measurement noise, have been shown to remain bounded with the discussed approaches or even to tend to zero. However, for practical purposes, analyzing robustness under measurement noise is of paramount importance. To our best knowledge, a formal analysis of the sensitivity to the noise is missing in the prescribed-time literature based on TVGs. 

Hence, in this note, we analyze the sensitivity to measurement noise of a class of prescribed-time controllers and differentiator algorithms characterized by what we call an \emph{absolute deadline}.
The defining property of systems with such an absolute deadline is that the convergence-time bound stays the same for every trajectory, rather than shifting along with the \textit{initial time instant} as it would be the case for a time-invariant system with fixed-time convergence.
We show that many existing prescribed-time algorithms based on TVGs exhibit an absolute deadline. Furthermore, we prove some significant \emph{inherent} performance limitations and lack of robustness to measurement noise appearing arbitrarily close to such an absolute deadline. In particular, for nonscalar systems, we show that arbitrarily small noise may result in arbitrarily large control or observation errors at the absolute deadline and can also lead to diverging trajectories.
We also discuss why some popular workarounds have important drawbacks, to motivate future works to further analyze the sensitivity to measurement noise and the suggested workarounds.

\noindent \textbf{Notation:}
Boldface lowercase and capital letters denote vectors and matrices, respectively. $\mathbb{R}$ is the set of real numbers.
Given a vector $\mathbf{v}\in\mathbb{R}^{n}$, $\|\mathbf{v}\|=\sqrt{\mathbf{v}
^T\mathbf{v}}$, where $\mathbf{v}
^T$ is the transpose of $\mathbf{v}$. Given a scalar $v\in\mathbb{R}$, $|v|$ represents its absolute value. The $i$-th element of a vector $\mathbf{x}\in\mathbb{R}^n$ is denoted by $x_i$ and $\|\mathbf{x}\|_\infty=\max_{i}|x_i|$. One-sided limits of a function $f$ at a time instant $T$ from below are written as $\lim_{t \to T^{-}} f(t)$, $\limsup_{t \to T^{-}} f(t)$.
In the formal proofs, the convention $\operatorname{sign}(0) = 1$ is used.

\section{Preliminaries and definitions}
\label{Sec:Prelim}
In this work, we study systems of the form
\begin{equation}
\label{eq:system}
    \dot{\mathbf{x}} = \mathbf{f}(t,\mathbf{x}, \bm{\eta}, \nd)
\end{equation}
defined for $t\in[0,T)$, $T>0$.
Therein, $\bm{\eta}(t) \in \mathbb{R}^n$ is a time-varying noise, and the disturbance $\nd(t)\in\mathbb{R}$ and $\mathbf{f}:[0,T)\times\mathbb{R}^n\times\mathbb{R}^n\times\mathbb{R}\to \mathbb{R}^n$ are assumed to be such that system \eqref{eq:system} has a unique Filippov solution $\mathbf{x}(t)\in\mathbb{R}^n$ defined on $t\in[0,T)$ \citep{Filippov1988DifferentialSides} for $\bm{\eta}(t)\equiv\bm{0}$.
Moreover, we assume that both $d$ and $\bm{\eta}$ are Lebesgue measurable, and that the noise $\bm{\eta}$ is uniformly bounded as $\|\bm{\eta}(t)\| \le \bar \eta$ for some $\bar \eta$ and all $t \in [0,T]$.

\begin{defn}
[{Uniform Lyapunov Stability\protect\footnotemark}]
\label{def:stability}
\footnotetext{Notice that uniform Lyapunov stability is a property that is usually defined on the time interval $[0, \infty)$, see~\citep[Definition~4.2]{Khalil2002NonlinearSystems}. For simplicity in the terminology, we define uniform Lyapunov stability restricted to the interval of interest $[0,T)$.}
Given $T>0$, we say that the origin of system~\eqref{eq:system} is \emph{uniformly Lyapunov stable} on $[0,T)$ if, for every $\epsilon > 0$, there exists $\delta > 0$ such that for all $s\in [0,T)$,
$\|\mathbf{x}(s)\| \le \delta$ implies $\|\mathbf{x}(t)\| \le \epsilon$ for all $t\in [s,T)$%
\end{defn}

In particular, we are interested in studying systems \eqref{eq:system} that satisfy the following property:

\begin{defn}[Absolute Deadline]
\label{def:absolute}
Given $T>0$, we say that system \eqref{eq:system} has an \emph{absolute deadline} at $t=T$ if, for any initial time instant $s\in [0,T)$ and $\bm{\xi}\in\mathbb{R}^n$, every solution $\mathbf{x} : [s,T) \to \mathbb{R}^n$ of \eqref{eq:system} with $\bm{\eta} = \bm{0}$ and $\mathbf{x}(s) = \bm{\xi}$ satisfies 
$\lim_{t\to T^-}\mathbf{x}(t)=\bm{0}$.
\end{defn}

Note that although its definition involves trajectories with different initial time instants, the absolute deadline is a mathematical property of system \eqref{eq:system} that is unrelated to the actual time instant when a controller or differentiator is switched on, commonly called $t_0$ in the literature, cf. \citet{Holloway2019}. 
The definition of absolute deadline does not require system \eqref{eq:system} to be undefined for $t\geq T$. Therefore, this definition and all further results apply also in the case of systems defined on the unbounded time interval $[0,\infty)$.

In this note, we focus on the subclass of prescribed-time algorithms  exhibiting an absolute deadline (prescribed-time algorithms ensure convergence before a user-defined time). However, note that not every prescribed-time algorithm exhibits an absolute deadline, such is the case of the time-invariant algorithm in~\citep{Seeber2021RobustTime}, or the time-varying algorithm with uniformly bounded TVG in~\cite{Aldana-Lopez2022AGains}. Two specific structures of system \eqref{eq:system} are considered that are particularly relevant in the context of either control or observation.
For control, the form
\begin{equation}
\label{eq:control}
    \begin{aligned}
    \dot{\nx}_{i} &= \nx_{i+1}, \quad i\in\{1,\dots,n-1\} \\
    \dot{\nx}_n &= v(t,\mathbf{x}+\bm{\eta}) + \nd(t)
    \end{aligned}
\end{equation}
with $n \ge 2$ and $v : [0,T) \times \mathbb{R}^n \to \mathbb{R}$ is considered.
Such a system is obtained as a closed loop when applying a TVG-based control law $v(t,\x)$ to a perturbed integrator chain.
For observation, a system of the form
\begin{equation}
\label{eq:diff}
    \begin{aligned}
    \dot{\nx}_{i} = \nx_{i+1} &+ \phi_i(t,\nx_1+\eta_1), \quad i\in\{1,\dots,n-1\} \\
    \dot{\nx}_n = \nd(t)&+\phi_n(t,\nx_1+\eta_1)
    \end{aligned}
\end{equation}
with $n \ge 2$, $\phi_1, \ldots, \phi_n : [0,T) \times \mathbb{R} \to \RR$, and measurement noise $\eta_1$ with $|\eta_1(t)| \le \bar \eta$.
Such a system models error dynamics when constructing a differentiator, i.e., an observer for the state of a perturbed integrator chain.

For example, consider the system with $n = 2$ and $T = 1$:
\ifx\zeropert\undefined%
\begin{equation}
\begin{aligned}
\label{eq:example_sys}
    &\dot{\nx}_1 = \nx_2, \ \ \dot{\nx}_2 = -\tfrac{6}{(1-t)^2}\nx_1-\tfrac{4}{1-t}\nx_2 + d(t)\\
    \end{aligned}
\end{equation}
with $d(t)=(1-t)^2$. Given an initial condition $[\xi_1,\xi_2]^T$ at $t=s$, the unique solution to \eqref{eq:example_sys} can be written as
\begin{align}
\label{eq:example_sol}
     \nx_1(t)&=\left( \tfrac{3(1-t)^2}{(1-s)^2}-\tfrac{2(1-t)^3}{(1-s)^3} \right)\xi_1 + \left(\tfrac{(1-t)^2}{(1-s)} - \tfrac{(1-t)^3}{(1-s)^2}\right)\xi_2 \displaybreak[0]\nonumber\\
     & + \tfrac{1}{2}(s-t)^2(1-t)^2\nonumber\displaybreak[0]\\
     \nx_2(t) &= \left( \tfrac{6(1-t)^2}{(1-s)^3}-\tfrac{6(1-t)}{(1-s)^2} \right)\xi_1 + \left(\tfrac{3(1-t)^2}{(1-s)^2} - \tfrac{2(1-t)}{(1-s)}\right)\xi_2 \nonumber\\
     & - (s-t)^2(1-t) - (s-t)(1-t)^2
\end{align}%
\else%
\begin{equation}
\begin{aligned}
\label{eq:example_sys}
    &\dot{\nx}_1 = \nx_2, \ \ \dot{\nx}_2=v(t,\mathbf{x}) = -\tfrac{6}{(1-t)^2}\nx_1-\tfrac{4}{1-t}\nx_2\\
    \end{aligned}
\end{equation}
All trajectories of this linear time-varying system may be written as
\begin{equation}
\begin{aligned}
\label{eq:example_sol}
     &\nx_1(t)=\left( \tfrac{3(1-t)^2}{(1-s)^2}-\tfrac{2(1-t)^3}{(1-s)^3} \right)\xi_1 + \left(\tfrac{(1-t)^2}{(1-s)} - \tfrac{(1-t)^3}{(1-s)^2}\right)\xi_2 \\
     &\nx_2(t) = \left( \tfrac{6(1-t)^2}{(1-s)^3}-\tfrac{6(1-t)}{(1-s)^2} \right)\xi_1 + \left(\tfrac{3(1-t)^2}{(1-s)^2} - \tfrac{2(1-t)}{(1-s)}\right)\xi_2
    \end{aligned}%
\end{equation}%
\fi%
for any $s \in [0,1)$ and $\xi_1,\xi_2 \in \mathbb{R}$, and hence satisfies
$\lim_{t\to 1^{-}}\x(t) = \bm{0}$.
Thus, system \eqref{eq:example_sys} has an absolute deadline at $t=T=1$.
The following proposition shows that every time-varying linear system with a fixed (prescribed) convergence time has an absolute deadline.

\begin{prop}
\label{prop:abs_deadline}
Consider system \eqref{eq:system} with $\mathbf{f}(t,\mathbf{x},\bm{0},\nd) =\mathbf{A}(t)\mathbf{x} + \mathbf{b}(t) d$ with $\mathbf{A}:[0,T)\to\mathbb{R}^{n\times n}$ and $\mathbf{b}: [0,T) \to \mathbb{R}^n$ continuous. Suppose that every solution of \eqref{eq:system} with $\bm{\eta} = \bm{0}$ starting at time $t = 0$ satisfies $\lim_{t\to T^{-}}\x(t) = \mathbf{0}$.
Then, \eqref{eq:system} has an absolute deadline\footnote{The same conclusion holds for every system defined by $\mathbf{f}(t,\mathbf{x},\bm{0},d(t)) =: \mathbf{g}(t,\mathbf{x})$ globally Lipschitz in $\mathbf{x}$, uniformly over $t$ in every compact subinterval of $[0,T)$.} at $T$.
\end{prop}

It follows from Proposition~\ref{prop:abs_deadline}, that some algorithms proposed in the literature induce an absolute deadline, e.g., \cite{Song2017Time-varyingTime,Holloway2019,Song2019Time-varyingTime}. In other cases, existence of Filippov solutions with an absolute deadline can be shown through a time-scaling argument, e.g., \cite{Pal2020DesignTime,Aldana-Lopez2021AGains,Orlov2022Prescribed-TimeGains,Tran2020Finite-timeApproach}. These algorithms have been used for control in \cite{Song2017Time-varyingTime,Gomez2020RNC,Pal2020DesignTime,Song2019Time-varyingTime}, for a system \eqref{eq:system} with $n\geq 2$ of the form \eqref{eq:control}.
Usually, the disturbance $\nd(t)$ therein is restricted to a class of admissible functions, e.g. measurable signals bounded by a constant $L>0$ for all $t\geq 0$.

Similarly, systems with an absolute deadline have been used for differentiation by \cite{Holloway2019,Aldana-Lopez2021AGains,Orlov2022Prescribed-TimeGains} by studying error systems of the form \eqref{eq:diff}.
For example, take the system from Example 1 in \cite{Holloway2019}:
\begin{equation}
\label{eq:holloway_example}
\begin{array}{ll}
    \dot x_1 = &- \left(\ell_1 + \tfrac{6}{T-t}\right) x_1+x_2\\
    \dot x_2 = &-\left(\ell_2 + \tfrac{3\ell_1}{T-t} + \tfrac{6}{(T-t)^2}\right) x_1
\end{array}
\end{equation}
which is of the form \eqref{eq:diff} with $n=2, d(t)=0$ and arbitrary $T,\ell_1,\ell_2>0$. It was shown by \cite{Holloway2019} that for any $\mathbf{x}(0)=\bm{\xi}$, \eqref{eq:holloway_example} satisfies $\lim_{t\to T^{-}}\mathbf{x}(t)=\bm{0}$. Thus, \eqref{eq:holloway_example} has an absolute deadline at $t=T$ by virtue of Proposition \ref{prop:abs_deadline}.
With similar arguments being applicable to the linear time-varying controllers by \cite{Song2017Time-varyingTime,Holloway2019,Song2019Time-varyingTime} discussed above and the mentioned time-scaling argument being applicable to other approaches such as  \citep{Aldana-Lopez2021AGains,Orlov2022Prescribed-TimeGains}, one can see that a significant number of prescribed-time algorithms exhibit an absolute deadline.

\section{Robustness and performance limitations}
\label{Sec:Main}

Despite the benefits of algorithms equipped with an absolute deadline, these systems have inherent limitations in terms of stability, robustness, and practical feasibility.

\subsection{Controllers for integrator chains}
\label{SubSec:Controller}

The following theorem establishes a set of consequences of systems of the form \eqref{eq:control} with $n\ge 2$ and an absolute deadline, i.e., perturbed integrator chains under prescribed-time control.

\begin{thm}
\label{thm:controller}
Let $n \ge 2$, $T > 0$ and suppose that system \eqref{eq:control} has an absolute deadline at $T$. Then,
\begin{enumerate}[label={\normalfont\textbf{\roman*})}]
    \item\label{itm:instability} For all $\bar{\eta} > 0$ and $\mathbf{x}_0 \in \mathbb{R}^n$
    there exists a piecewise continuous and bounded noise $\bm{\eta}:[0,T)\to\mathbb{R}^n$ with countably many discontinuities and $\|\bm{\eta}(t)\|\leq \bar{\eta}, \forall t\in[0,T]$ such that the solution $\mathbf{x}(\cdot)$ with $\mathbf{x}(0)=\mathbf{x}_0$ satisfies $\limsup_{t\to T^-}\|\mathbf{x}(t)\| = \infty$.
    \item\label{itm:accuracy}
    For all $\epsilon > 0$, if there exists a continuous $\bm{\xi}_\epsilon:[0,T]\to\mathbb{R}^{n-1}$ satisfying $\bm{\xi}_\epsilon(T)=0$ and $\limsup_{t \to T^-} |v(t,[\bm{\xi}_\epsilon(t)^T,-2\epsilon]^T)| < \infty$, then, for all $\bar \eta > 0$ and all $\bm{x}_0 \in \mathbb{R}^n$ there exists a piecewise continuous noise $\bm{\eta}:[0,T)\to\mathbb{R}^n$ with two discontinuities and $\|\bm{\eta}(t)\| \le \bar \eta$ such that the solution $\mathbf{x}(\cdot)$ with $\mathbf{x}(0)=\mathbf{x}_0$ satisfies $\lVert\lim_{t\to T^{-}} \mathbf{x}(t)\rVert \ge \epsilon$.
    \item\label{itm:bounded} For all $\delta>0$, $\sup_{\|\mathbf{x}\|_{\infty}\le \delta, t\in[0,T)} |v(t,\mathbf{x})| = \infty$.
    \item\label{itm:c_unif} The origin of~\eqref{eq:control} is not uniformly Lyapunov stable.
\end{enumerate}
\end{thm}

\begin{rem}
Theorem~\ref{thm:controller}- \ref{itm:instability} and~\ref{itm:accuracy} show that arbitrarily small noises can lead to trajectories with arbitrarily large or even diverging control error at the deadline $T$.
As a consequence, arbitrarily bad tracking performance on the interval $[0,T)$ is obtained even if such noise is removed in a vicinity of the deadline $T$.
Theorem~\ref{thm:controller}-\ref{itm:bounded} and~\ref{itm:c_unif} show the main reasons for this lack of robustness: unboundedness of the controller $v(t,\mathbf{x})$ in $t$, which is consistent with literature on prescribed-time control based on TVGs, and more importantly, lack of uniform Lyapunov stability.
\end{rem}

\begin{rem}
\label{rem:lyapunov}
Theorem~\ref{thm:controller}-\ref{itm:c_unif} implies lack of uniform Lyapunov stability also in the classical sense of \cite[Definition~4.2]{Khalil2002NonlinearSystems}, if system \eqref{eq:system} is defined on the unbounded time interval, i.e., for $t \in [0, \infty)$.
\end{rem}

\begin{rem}
Note that the additional condition in Theorem \ref{thm:controller}-\ref{itm:accuracy} is very mild.
Often it is possible to achieve even $v(t, \bm{\xi}_{\epsilon}(t)^T, -2\epsilon]^T) = 0$.
In system \eqref{eq:example_sys}, this is achieved with $\xi_\epsilon(t) = \epsilon(4/3)(1-t)$, which satisfies $\xi_\epsilon(1) = 0$.
\end{rem}

\subsection{Differentiators}
\label{SubSec:Differ}

Similarly to the previous section, the following theorem establishes a set of consequences of systems of the form \eqref{eq:diff} with $n\ge 2$ and an absolute deadline, i.e., differentiation error dynamics with prescribed-time convergence.
\begin{thm}
\label{thm:differentiator}
Let $n \ge 2$, $T > 0$ and suppose that system \eqref{eq:diff} has an absolute deadline at $T$. Then,
\begin{enumerate}[label={\normalfont\textbf{\roman*})}]
    \item\label{itm:existence} For all $\bar{\eta} > 0$ and $\mathbf{x}_0 \in \RR^n$, there exists a locally Lipschitz continuous noise $\eta_1 : [0,T) \to [-\bar\eta,\bar\eta]$ such that the corresponding solution $\mathbf{x}(\cdot)$ with $\mathbf{x}(0)=\mathbf{x}_0$ satisfies $\limsup_{t\to T^-}\|\mathbf{x}(t)\| = \infty$.
    \item\label{itm:d_accuracy} For all $\bar \eta > 0$ and $\epsilon > 0$, there exists a Lipschitz continuous noise $\eta_1 : [0,T) \to [-\bar\eta,\bar\eta]$, such that all corresponding solutions satisfy $\lVert\lim_{t\to T^-} \x(t)\rVert \ge \epsilon$.
    \item\label{itm:d_bounded} For all $\delta>0$, $\sup_{|{\nx}_1| \le \delta, t\in[0,T)} |\bm{\phi}_i(t,{\nx}_1)| = \infty$ for some $i\in\{1,\dots,n\}$.
\end{enumerate}
\end{thm}

\begin{rem}
Note that unlike Theorem \ref{thm:controller}, lack of uniform Lyapunov stability is not shown in Theorem \ref{thm:differentiator}.
Nevertheless, an equivalent set of consequences to those in Theorem \ref{thm:controller} is also shown in this case.
In fact, Theorem~\ref{thm:differentiator}-\ref{itm:d_accuracy} is even stronger than Theorem~\ref{thm:controller}-\ref{itm:d_accuracy}, because an arbitrarily small noise signal yields arbitrarily bad performance \emph{for all initial conditions} in that case.
\end{rem}

\begin{rem}
\label{Rem:Robust}
    A robust exact differentiator, in the sense of~\cite{Levant1998RobustTechnique,seeber2023}, is one which differentiates noise-free signals exactly after a finite (possibly prescribed) time, and whose behavior under bounded noise tends uniformly to the behavior in the absence of noise as the noise bound tends to zero. An important corollary of Theorem~\ref{thm:differentiator} is that \emph{{a robust exact differentiator with an absolute deadline cannot exist}}. %
\end{rem}

\section{Discussion}
\label{Discussion}

In the prescribed-time literature based on TVG it is often acknowledged that it is problematic, when measurement noise is present, to have TVG that tends to infinity, see e.g.,~Section~3.2 in \cite{Song2017Time-varyingTime} and Section~2.A in \cite{Holloway2019}. To avoid this problem, some workarounds have been suggested in the literature.

A common workaround proposed by~\cite{Song2017Time-varyingTime,Holloway2019} is to switch off the algorithm at a time $\toff$ before the absolute deadline $T$, thus maintaining the TVG uniformly bounded. With such a workaround, an absolute deadline is no longer present, eliminating the lack of robustness exposed above,
but at the same time also the convergence to zero in prescribed time.
As a result, the error then does not reach zero at time $t\sbrm{stop}$, and for the case of linear time-varying systems with bounded dynamic matrix on $[0, \toff]$, the remaining error grows linearly with the initial condition.
Moreover, our results show that the system becomes more sensitive to measurement noise as $\toff$ approaches $T$.

Another workaround proposed by~\cite{Song2017Time-varyingTime} is to switch off the algorithm when the error trajectory enters a desired deadzone on the error.
A particular case of this workaround is essentially used in \cite{Orlov2022Prescribed-TimeGains}, with a deadzone of zero width.
However, with such an approach, due to the nature of the algorithms in~\cite{Song2017Time-varyingTime,Song2019Time-varyingTime,Holloway2019,Orlov2022Prescribed-TimeGains}, unperturbed trajectories with a large initial condition will enter such deadzone arbitrarily close to $T$ with arbitrarily large TVG.
The presence of additional arbitrarily small noise may furthermore prevent the trajectory from entering the deadzone at all, thus lacking robustness despite the workaround. Indeed, in the case of the  zero-width deadzone workaround, used e.g., in~\cite{Orlov2022TimeDesigns,Orlov2022Prescribed-TimeGains,VerdesKairuz2022RobustGains}, it can be shown, using a time-scale transformation argument, that the workaround does not eliminate the absolute deadline property.%

\section{Illustrative Example}
\label{sec:example}
Recall system \eqref{eq:example_sys} under $d(t) = (1-t)^2$, which was shown above to exhibit an absolute deadline at $t=T=1$. Consider a noise signal $\bm{\eta}(t) = [\eta(t), 0]^T$ with $\eta(t)=\bar{\eta}\sqrt{1-t}$  satisfying $\|\bm{\eta}(t)\|=|\eta(t)|\leq\bar{\eta}, \forall t\in[0,1)$ such that system \eqref{eq:example_sys} becomes:
\begin{equation}
\begin{aligned}
\label{eq:example_sys2}
    &\dot{\nx}_1 = \nx_2, \ \ \dot{\nx}_2 = -\tfrac{6}{(1-t)^2}(\nx_1+ \eta(t))-\tfrac{4}{1-t}\nx_2 + d(t)\\
    \end{aligned}
\end{equation}
It can be verified that $\mathbf{x}(t) = [x_1(t), x_2(t)]^T$ with
\begin{equation}
\label{eq:example_sys2_sol}
\begin{aligned}
x_1(t) =& (3(1-t)^2 - 2(1-t)^3)\xi_1 \\
&+ ((1-t)^2 - (1-t)^3)\xi_2 + \frac{1}{2}t^2(1-t)^2 \\
&+ \frac{4\bar{\eta}}{5}\bigg( 5(1-t)^2- 3(1-t)^3 -2\sqrt{1-t} \bigg) \\
x_2(t) =& (6(1-t)^2 - 6(1-t))\xi_1 - t^2(1-t) \\
&+ (3(1-t)^2 - 2(1-t))\xi_2 + t(1-t)^2 \\
&+ \frac{4\bar{\eta}}{5}\bigg( 9(1-t)^2 - 10(1-t) \bigg)+ \frac{4\bar{\eta}}{5\sqrt{1-t}} 
\end{aligned}
\end{equation}
is the unique solution of \eqref{eq:example_sys2} for $t\in[0,1)$ with $\mathbf{x}(0)= [\xi_1, \xi_2]^T$. In case $\bar \eta = 0$, $\lim_{t\to T^{-}}\mathbf{x}(t) = \bm{0}$, verifying that the system has an absolute deadline at $t = T$. As already discussed, $\mathbf{x}(\toff)$ is an unbounded function of the initial condition $\mathbf{x}(0)$ and hence switching off the algorithm at a time $\toff<T$ is not sufficient for convergence.

\begin{figure}
    \centering
    \def\svgwidth{0.45\textwidth}
    \begingroup%
  \makeatletter%
  \providecommand\color[2][]{%
    \errmessage{(Inkscape) Color is used for the text in Inkscape, but the package 'color.sty' is not loaded}%
    \renewcommand\color[2][]{}%
  }%
  \providecommand\transparent[1]{%
    \errmessage{(Inkscape) Transparency is used (non-zero) for the text in Inkscape, but the package 'transparent.sty' is not loaded}%
    \renewcommand\transparent[1]{}%
  }%
  \providecommand\rotatebox[2]{#2}%
  \newcommand*\fsize{\dimexpr\f@size pt\relax}%
  \newcommand*\lineheight[1]{\fontsize{\fsize}{#1\fsize}\selectfont}%
  \ifx\svgwidth\undefined%
    \setlength{\unitlength}{207.45774841bp}%
    \ifx\svgscale\undefined%
      \relax%
    \else%
      \setlength{\unitlength}{\unitlength * \real{\svgscale}}%
    \fi%
  \else%
    \setlength{\unitlength}{\svgwidth}%
  \fi%
  \global\let\svgwidth\undefined%
  \global\let\svgscale\undefined%
  \makeatother%
  \begin{picture}(1,0.49067141)%
    \lineheight{1}%
    \setlength\tabcolsep{0pt}%
    \put(0,0){\includegraphics[width=\unitlength,page=1]{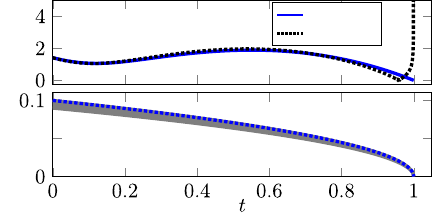}}%
    \put(0.72350702,0.44485705){\makebox(0,0)[lt]{\lineheight{1.25}\smash{\begin{tabular}[t]{l}\small$\bar{\eta}=0$\end{tabular}}}}%
    \put(0.72350702,0.39830879){\makebox(0,0)[lt]{\lineheight{1.25}\smash{\begin{tabular}[t]{l}$\bar{\eta}=0.1$\end{tabular}}}}%
    \put(0.03702874,0.13){\rotatebox{90}{\makebox(0,0)[lt]{\lineheight{1.25}\smash{\begin{tabular}[t]{l}$\eta(t)$\end{tabular}}}}}%
    \put(0.03702874,0.32609479){\rotatebox{90}{\makebox(0,0)[lt]{\lineheight{1.25}\smash{\begin{tabular}[t]{l}$\|\mathbf{x}(t)\|$\end{tabular}}}}}%
  \end{picture}%
\endgroup%
    \caption{Above: Solutions to \eqref{eq:example_sys2} with explicit expression given in \eqref{eq:example_sys2_sol}, with $\xi_1=\xi_2=1$, and $\bar{\eta}=0$ and $\bar{\eta}=0.1$, respectively. Below: The dashed blue line represents the noise signal $\eta(t)=\bar{\eta}\sqrt{1-t}$. Any noise signal contained in the gray region $\bar{\eta}\sqrt{1-t}[0.9,1]$ causes a divergent trajectory.}
    \label{fig:ExampleNoise}
\end{figure}
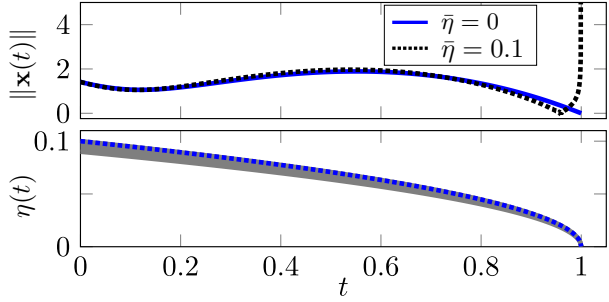

In the case $\bar{\eta}>0$, regardless of how small it is, the last term of $x_2(t)$ in~\eqref{eq:example_sys2_sol} is divergent at the deadline and thus $\limsup_{t\to T^-}\|\mathbf{x}(t)\| = \infty$. The trajectories $\mathbf{x}(t)$ for $\bar{\eta}=0$ and $\bar{\eta}=0.1$, respectively, with $\xi_1=\xi_2=1$ are illustrated in the first plot of Fig.~\ref{fig:ExampleNoise}. 
Moreover, it can be verified that any noise satisfying $\eta(t) \in (1-t)^\alpha \bar{\eta}[\beta,1]$  with $\frac{6-3\alpha}{6-2\alpha}<\beta\leq 1, \alpha\in(0,1)$ would also produce a divergent trajectory. The second plot of Fig.~\ref{fig:ExampleNoise}, illustrates this region in gray for $\alpha=\frac{1}{2}, \beta=0.9, \bar \eta = 0.1$, together with the noise $\eta(t)=0.1\sqrt{1-t}$ in dashed line.

\section{Conclusion}
\label{Sec:Conclusion}
In this note, we analyzed the behavior under measurement noise of a class of non-scalar fixed-time algorithms characterized by an absolute deadline. This analysis exposes some inherent performance limitations and lack of robustness, mainly when noise appears arbitrarily close to the absolute deadline. We show, for instance, that an arbitrarily small noise signal may result in arbitrarily large errors at the absolute deadline and can also lead to divergence. In line with existing literature, our analysis focuses on controllers for integrator chains and differentiators. 
In future work, we consider extending the study to more general forms of control algorithms, such as controllers for nonlinear systems and systems in a strict feed-forward form exhibiting an absolute deadline.

\appendix 
\section{Appendix}
\subsection{Proof of Proposition \ref{prop:abs_deadline}}
\label{ap:abs_deadline}
Consider an arbitrary $\mathbf{x}(s)=\bm{\xi}$ with $s\in[0,1)$. Hence, due to the proposition's assumptions, the right-hand side of \eqref{eq:system} is continuous in $t\in[0,s]$ and there exists $L > 0$ such that $\|\mathbf{f}(t,\mathbf{x}_1,\mathbf{0},d) - \mathbf{f}(t,\mathbf{x}_2,\mathbf{0},d)\| \le L \|\mathbf{x}_1 - \mathbf{x}_2\|$ for all $t \in [0, s]$ and all $\mathbf{x}_1, \mathbf{x}_2 \in \mathbb{R}^n$. Therefore, the Cauchy–Lipschitz existence theorem ensures that the solution $\mathbf{x}(t)$ can be continued backwards in time towards $\mathbf{x}(0) = \bm{\xi}'$. Hence, by the proposition assumptions, it follows that $\lim_{t\to T^{-}}\mathbf{x}(t)=\bm{0}$.

\subsection{Proof of Theorem \ref{thm:controller}}
First, we show the following auxiliary lemma:
\begin{lemma}
\label{lem:traj}
If there exists $T>0$ such that \eqref{eq:control} with $n\geq 2$ has an absolute deadline at $t=T$ then, for any pair $\delta,\epsilon>0$ there exists $s\in(0,T)$ such that if $|\nx_1(s')|\geq \delta$ for any $s'\in[s,T)$, then $\|\mathbf{x}(t)\|>\epsilon$ for some $t\in(s',T)$. 
\end{lemma}

\begin{pf} Given $\delta,\epsilon>0$ choose $s = T -\delta/\epsilon'$ for any $\epsilon'>\max(\epsilon,\delta)$ such that $s\in[0,T)$. Note that $\lim_{t\to T^{-}}\nx_1(t)=0$ from the absolute deadline property. Assume $|\nx_1(s')|\geq \delta$ for arbitrary $s'\in[s,T)$. By virtue of being a solution, $\nx_2(\bullet)$ is absolutely continuous and since $\dot{\nx}_1 = \nx_2$, then $\nx_1$ is differentiable everywhere in $(s,T)$. Hence, by the mean value theorem, there must exist $t\in(s',T)$ such that $\dot{\nx}_1(t)=\nx_2(t) = \frac{-\nx_1(s')}{T-s'}$. Therefore,
$
\|\mathbf{x}(t)\|\geq |\nx_2(t)| = \frac{|\nx_1(s')|}{T-s'}\geq \frac{|\nx_1(s')|}{T-s}\geq  \frac{\delta}{\delta/\epsilon'} = \epsilon'>\epsilon.
$
\end{pf}

\textbf{Item \ref{itm:instability}}: 
Consider strictly increasing sequences $\{t_k\}_{k=0}^\infty$, $ \{t_k'\}_{k=0}^\infty$ with $t_0=t_0'=0$ and $\{\epsilon_k\}_{k=0}^\infty$ with $\lim_{k\to\infty}\epsilon_k=\infty$. Given $\delta\in(0,\bar{\eta})$, we construct a noise $\bm{\eta}(t)=\eta_1(t)\mathbf{b}_1$ with $\eta_1(t) = \delta\text{sign}(\nx_1(t_k)), \forall t\in[t_k, t_{k+1})$ which is possible due to causality of \eqref{eq:control}. Note that $\|\bm{\eta}(t)\|\leq \bar{\eta}, \forall t\in[0, \lim_{k\to\infty} t_k)$. Now, we construct the rest of $\{t_k\}_{k=1}^\infty, \{t_k'\}_{k=1}^\infty$ as follows. Given the pair $\delta,\epsilon'_k>0$ with $\epsilon'_k=\epsilon_k+\bar{\eta}$, use $s>0$ as in Lemma~\ref{lem:traj} to define any $t_k\in( \max(s,t_{k-1}, t_{k-1}',T-1/\epsilon_k), T)$ picked such that in the case of $\bar{\eta}=0$, $|\nx_1(t_k)|\geq \delta$ implies $\|\mathbf{x}(t_k')\|>\epsilon_k$ for some $t_k'\in(t_k,T)$ as from Lemma~\ref{lem:traj}. Note that $t_k\geq T-1/\epsilon_k$ such that both $\lim_{k\to\infty} t_k = \lim_{k\to\infty} t_k' = T$. Given this construction of $\bm{\eta}(t)$, analyze the interval $[t_k, t_{k+1})$. Let $\mathbf{z}(t) = \mathbf{x}(t)+\bm{\eta}(t)$. For this interval, we have $ \dot{\nz}_i = \nz_{i+1}$ for $i=1,\ldots,n-1$ and $\dot{\nz}_n = v(t,\mathbf{z})+\nd(t)$,
which is precisely \eqref{eq:diff} with $\bar{\eta}=0$. Moreover, note that 
$
|\nz_1(t_k)| = |\nx_1(t_k) + \delta \text{sign}(\nx_1(t_k))| = |\nx_1(t_k)|+\delta\geq \delta
$
Thus, $\|\mathbf{z}(t_k')\|>\epsilon'_k$ by Lemma~\ref{lem:traj} and the definition of $t_k'$ which implies
$
\epsilon_k+\bar{\eta}=\epsilon'_k<\|\mathbf{z}(t_k')\|=\|\mathbf{x}(t_k')+\bm{\eta}(t_k')\|\leq \|\mathbf{x}(t_k')\|+\bar{\eta}
$.
Thus, $\|\mathbf{x}(t_k')\|>\epsilon_k$. As a consequence,
$
\limsup_{t\to T^-}\|\mathbf{x}(t)\|\geq \limsup_{k\to \infty}\|\mathbf{x}(t_k')\|>\sup_{t\to\infty}\epsilon_k = \infty
$
concluding the proof for this item.

\textbf{Item \ref{itm:accuracy}}: Let $\bar{\eta}'=\bar{\eta}/\sqrt{n}$, $\bm{\xi}' = [\bm{\xi}_{\epsilon}^T, 0]^T$, $\bm{q} = [\bm{0}^T, -2\epsilon]^T$ and $\Psi_{n+1}(t) = v(t, \bm{\xi}'(t)+ \bm{q})$.
By assumption, $s > \max(T-{\bar{\eta}'}/{(12\epsilon)},T-1/2)$ exists such that $(T - s)|\Psi_{n+1}(t)| \le \min(\epsilon, \bar\eta'/2)$ and  $\|{\bm{\xi}_{\epsilon}(t)}\|_{\infty} \le \bar \eta'/2$ for all $t \in [s, T)$.
Define functions $\Psi_i : [s,T) \to \RR$ for $i = 1, \ldots, n$ recursively via $\dot \Psi_{i}(t) = \Psi_{i+1}(t)$ with $\Psi_{n}(s) = -2\epsilon$ and arbitrary $\Psi_{i}(s) \in [-\bar \eta'/4,\bar \eta'/4]$ for $i < n$.
From the bound on $|\Psi_{n+1}(t)|$, obtain $\Psi_{n}(t) \in [-3\epsilon, -\epsilon]$ and $|\Psi_n(t) - \Psi_n(s)| \le \bar \eta'/2$.
From $t - s \le \bar\eta'/(12 \epsilon)$, then $|\Psi_{n-1}(t)| \le \bar \eta'/2$, and using $t - s \le 1/2$ yields $|\Psi_{i}(t)| \le \bar \eta'/2$ for all $i = 1, \ldots, n-1$.
With $\bm{\Psi} = [\Psi_1, \ldots, \Psi_n]^T$, then $\|\bm{\Psi}(t) - \bm{q}\|_{\infty} \le \bar\eta'/2$.
Now, define the noise as $\bm{\eta}(t) = \bm{\xi}'(t) + \bm{q} - \bm{\Psi}(t)$, which satisfies $\|\bm{\eta}(t)\|_{\infty} \le \|\bm{\xi}'(t)\|_{\infty} + \bar \eta'/2 \le \bar \eta'$ and hence $\|\bm{\eta}(t)\| \le \bar \eta' \sqrt{n} = \bar \eta$.
Then, $\x(t) = \bm{\Psi}(t)$ is a solution of \eqref{eq:control} on $[s,T]$, because $\dot \nx_i(t) = \nx_{i+1}(t)$ for $i < n$ and $\dot \nx_n(t) = \Psi_{n+1}(t) = v(t, \bm{\xi}'(t) + \bm{q}) = v(t, \x(t) + \bm{\eta}(t))$ by construction, with $\|\lim_{t\to T^{-}}\x(t)\| \ge |\lim_{t\to T^{-}}\Psi_n(t)| \ge \epsilon$.
To steer every initial condition $\mathbf{x}(0)$ to an $\mathbf{x}(s)$ of the required form, apply a constant noise $\bm{\eta}(t) = [\bar\eta'/8, 0, \ldots, 0]^T$ initially, leading to $\lVert \bm{x}(s_0) - \bm{\eta}(s_0)\rVert \le \min(\bar\eta'/16,\epsilon)$ for some $s_0 > \max(T-\bar\eta'/(32\epsilon),T-1/2)$, i.e., $x_1(s_0) \in [\bar\eta'/16,3\bar\eta'/16]$, $|x_i(s_0)|\le \bar\eta'/16$ for $i=2,\ldots,n$, and $|x_n(s_0)| \le \epsilon$.
Removing the noise, i.e., setting $\bm{\eta}(t) = \bm{0}$ starting at $t \ge s_0$, then yields $\x(s)$ of the required form (possibly with reversed sign) at $s = \inf \{ \sigma \ge s_0 : |x_n(\sigma)| \ge 2\epsilon\}$.

\textbf{Item \ref{itm:bounded}}: Assume $\sup_{\mathbf{x}\in[-\delta,\delta]^n, t\in[0,T]} |v(t,\mathbf{x})| = \epsilon$ for some $\epsilon\geq 0$. Consider $s=T-\delta/\epsilon'$ with $\epsilon'>\max(\epsilon,\delta)$ and an arbitrary initial condition $\mathbf{x}(s)$ for \eqref{eq:control} with $\bar{\eta}=0$ and $\mathbf{x}(s)\in[-\delta,\delta]^n$. The absolute deadline property $\lim_{t\to T^{-}}\mathbf{x}(t)=0$ implies the existence of $s'=\inf\{ t\in[s,T) : \mathbf{x}(t')\in[-\delta,\delta]^n , \forall t'\geq t \}$ and, by absolute continuity of $\mathbf{x}(t)$, $\nx_i(s')=\delta$ for some $i\in\{1,\dots,n\}$. Assume $i\neq n$; by the mean value theorem there then exists $t\in(s',T)$ with $\dot{\nx}_{i}(t) = \nx_{i+1}(t) = \frac{-\nx_i(s')}{T-s'}$. Therefore,
$
|\nx_{i+1}(t)| = \frac{|\nx_i(s')|}{T-s'}\geq \frac{|\nx_i(s')|}{T-s} = \frac{\delta}{\delta/\epsilon'} = \epsilon'>\delta
$
which is a contradiction of $\mathbf{x}(t)\in[-\delta,\delta]^n, \forall t\in[s',T)$. Hence,  $i=n$. The function $\nx_{n}$ is absolutely continuous and thus differentiable almost everywhere. Therefore, it can be shown that there must exist $t_1,t_2\in(s',T)$ such that $v(t_1,\mathbf{x}(t_1)) \le \frac{-\nx_{n}(s')}{T-s'}$ and $v(t_2,\mathbf{x}(t_2)) \ge \frac{-\nx_{n}(s')}{T-s'}$.
Therefore, there exists $t\in\{t_1,t_2\}$ so that $|v(t,\mathbf{x}(t))| \geq \frac{|\nx_{n}(s')|}{T-s}=\epsilon'>\epsilon$. The previous fact, in addition to $\mathbf{x}(t)\in[-\delta,\delta]^n$, contradicts the initial assumption.

\textbf{Item \ref{itm:c_unif}}: We will show that for any $\delta,\epsilon>0$, there exist $s,t$ with $0\leq s<t\leq T$ and a trajectory of \eqref{eq:control} which satisfies both $\|\mathbf{x}(s)\|\leq \delta$ and $\|\mathbf{x}(t)\|> \epsilon$, which implies that uniform Lyapunov stability does not hold. Given $\delta,\epsilon>0$, choose $s\in(0,T)$ as in Lemma~\ref{lem:traj} and a trajectory of \eqref{eq:control} passing through $\mathbf{x}(s) = \delta\mathbf{b}_1$ satisfying $\|\mathbf{x}(s)\|=|\nx_1(s)|=\delta$. Hence, Lemma~\ref{lem:traj} implies $\|\mathbf{x}(t)\|>\epsilon$ for some $t\in(s,T)$.

\subsection{Proof of Theorem \ref{thm:differentiator}}

\textbf{Item \ref{itm:existence}}: Consider strictly increasing sequences $\{t_k\}_{k=0}^{\infty}$, $\{\epsilon_k\}_{k=0}^{\infty}$ with $t_0=\epsilon_0 = 0$ and $\lim_{k\to\infty}\epsilon_k=\infty$. Now, let $\eta_1(t) = -(\bar{\eta}\text{sign}(\eta_1(t_k)) + \eta_1(t_k))(t-t_k)/(T-t_k) + \eta_1(t_k) $ for all $t\in[t_k,t_{k+1})$ and $\eta_1(0)=\bar{\eta}$, which is locally Lipschitz continuous. It can be verified that since $|{\eta}_1(0)|\leq \bar{\eta}$, then $|\eta_1(t)|\leq \bar{\eta}, \forall t\in[0,\sup \{t_k\}_{k=0}^\infty)$. Now, note that $|\dot{\eta}_1(t)|>\epsilon_0, \forall t\in[t_0,t_1)$ for arbitrary $t_1$. Hence, we construct the rest of the $t_k$ recursively as follows. First, assume $|\dot{\eta}_{1}(t)|=(\bar{\eta}+|\eta_1(t_k)|)/(T-t_k)>\epsilon_k, \forall t\in[t_k,t_{k+1})$. Then, given $\delta>0$ there exists $t'_{k+1}\in(t_k,T)$  such that for $t\in[t_{k+1},t_{k+2})$ we have $|\dot{\eta}_1(t)|=(\bar{\eta}+|\eta_1(t_{k+1})|)/(T-t_{k+1})>\epsilon_k+\delta$ for arbitrary $t_{k+1}\in(t_{k+1}',T)$ and $t_{k+2}\in(t_{k+1},T)$. Moreover, on the time interval $[t_k, t_{k+1})$, consider new state variables $\nz_1 = \nx_1 + \eta_1$, $\nz_2 = \nx_2 + \dot{\eta}_1$, and $\nz_{i} = \nx_i$ for $i > 2$ leading to
\begin{equation}
\label{eq:aux_sys}
    \begin{aligned}
    \dot{\nz}_i &= \nz_{i+1} + \phi_i(t,\nz_1), \quad i\in\{1,\dots,n-1\} \\
    \dot{\nz}_n &= \nd(t)+\phi_n(t,\nz_1)
    \end{aligned}
\end{equation}
Hence, given $\delta>0$ there exists $t_{k+1}''\in(t_k,T)$ such that $|\nz_2(t_{k+1})|\leq \delta$, equivalently $|\nx_2(t_{k+1})-\dot{\eta}_1(t_{k+1})|\leq \delta, \forall t_{k+1}\in(t_{k+1}'',T)$ by virtue of the absolute deadline property. Hence, choose $t_{k+1}\in( \max(t_{k+1}',t_{k+1}'',T-1/\epsilon_k),T)$ implying that $|\nx_2(t_{k+1})|>\epsilon_k$ and $\sup\{t_k\}_{k=0}^\infty=T$. Similarly to the proof of Theorem~\ref{thm:controller}-\ref{itm:instability}, it follows that $\limsup_{t\to T^-}\|\x(t)\|=\infty$.

\textbf{Item \ref{itm:d_accuracy}}: Let $s = T - 2 \bar \eta/\epsilon$ and define the noise $\eta_1(t)=-\bar{\eta}$ for $t\in[0,s)$ and $\eta_1(t)=-\bar \eta + (t - s)\epsilon$ for $t\in[s,T)$. Note that $\eta_1$ is Lipschitz continuous and satisfies $|\eta_1(t)| \le \bar \eta$ for all $t \in [0, T]$.
On the time interval $(s, T]$, consider new state variables $\nz_1 = \nx_1 + \eta_1$, $\nz_2 = \nx_2 + \epsilon$, and $\nz_{i} = \nx_i$ for $i > 2$.
Since $\dot \eta_1(t) = \epsilon$ for $t \in (s, T]$ this leads to dynamics of the form \eqref{eq:aux_sys}
on this interval, and consequently $\lim_{t\to T^{-}}\nz_2(t) = 0$ by virtue of the absolute deadline property.
Hence, $\lim_{t\to T^{-}}\nx_2(t) = -\epsilon$ and $\lVert \lim_{t\to T^-} \x(t) \rVert \ge |\lim_{t\to T^{-}}\nx_2(t)| = \epsilon$.

\textbf{Item \ref{itm:d_bounded}}: Assume that $\sup_{{x}_1\in[-\delta,\delta], t\in[0,T]} \|\bm{\phi}(t,{x}_1)\| = \epsilon\geq 0$, let $\epsilon'>\max(\delta,\epsilon+\delta)$ and define $s,s'>0$ and initial condition $\mathbf{x}(s)$ in the same way as in the proof of Theorem \ref{thm:controller}-\ref{itm:bounded}. Hence, there is $i\in\{1,\dots n\}$ with $\nx_i(s')=\delta$ and $\mathbf{x}(t)\in[-\delta,\delta]^n, \forall t\in[s',T)$. Assume $i\neq n$ such that by the mean value theorem and the absolute deadline property there exists $t\in[s',T)$ with 
$
|\nx_{i+1}(t) + \phi_i(t,\nx_1(t))| = \frac{|\nx_i(t)|}{T-s} \geq \epsilon' > \epsilon + \delta
$. 
But $|\nx_{i+1}(t) + \phi_i(t,\nx_1(t))|\leq |\nx_{i+1}(t)| + |\phi_i(t,\nx_1(t))|\leq |\phi_i(t,\nx_1)| + \delta$. Hence, $\|\bm{\phi}(t,\nx_1(t))\|\geq |\phi_i(t,\nx_1(t))|>\epsilon$ and $\nx_1(t)\in[-\delta,\delta]$ contradicting the initial assumption. The proof follows in a similar way when $i=n$.

\end{document}